\documentstyle[12pt,epsfig]{article}
\begin{document} \setlength{\baselineskip}{5mm}

\begin{center}

{\Large \bf Wigner's Spins, Feynman's Partons, and Their Common
Ground}\\[4mm]

{\large \bf Y. S. Kim} \\[4mm]

{\it Department of Physics, University of Maryland, College Park,
  Maryland 20742, U.S.A.}

\end{center}

\vspace{10mm}

\begin{abstract}
The connection between spin and symmetry was established by Wigner
in his 1939 paper on the Poincar\'e group. For a massive particle
at rest, the little group is $O(3)$ from which the concept of
spin emerges.  The little group for a massless particle is isomorphic
to the two-dimensional Euclidean group with one rotational and two
translational degrees of freedom.  The rotational degree corresponds to
the helicity, and the translational degrees to the gauge degree of
freedom.   The question then is whether these two different symmetries
can be united.  Another hard-pressing problem is Feynman's parton
picture which is valid only for hadrons moving with speed close to
that of light.  While the hadron at rest is believed to be a bound
state of quarks, the question arises whether the parton picture is
a Lorentz-boosted bound state of quarks.  We study these problems
within Einstein's framework in which the energy-momentum relations
for slow particles and fast particles are two different manifestations
one covariant entity.
\end{abstract}

\vspace{15mm}

This note is based on the lectures delivered at the Advanced Study
Institute on Symmetries and Spin (Praha, Czech Republic, July 2001).

\newpage

\section{Introduction}\label{intro}

Let us start with Einstein's energy-momentum relation.
If the momentum of a particle is much smaller than its mass, the
energy-momentum relation is $E = p^{2}/2m + mc^{2}$.  If the momentum
is much larger than the mass, the relation is $E = cp$.  These two
different relations can be combined into one covariant formula
$E = \sqrt{m^{2} + p^{2}}$.

There are other problems of similar nature.  Particles have internal
space-time variables.  Massive
particles have spins while massless particles have their helicities
and gauge degrees of freedom.  Can these symmetries be unified
like Einstein's energy-momentum relation?  These is another problem
in modern physics.  Hadrons are bound states of quarks.  However,
fast-moving hadrons are incoherent collections of partons, as Feynman
correctly observed in 1969~\cite{fey69}.  Can the quark model and the
parton model be combined into a single covariant entity?

We shall discuss these problems in this report.  First, let us review
Wigner's work.  In 1939~\cite{wig39}, Eugene Wigner constructed the
maximal subgroups of the Lorentz group whose transformations leave
the four-momentum of a given particle invariant.  These groups are
known as Wigner's little groups.
Thus, the transformations of the little groups change the internal
space-time variables of the particle, while leaving its four-momentum
invariant.  He observed that the little groups for massive and
massless particles are isomorphic to the three-dimensional rotation
group and the two-dimensional Euclidean group respectively.

If a massive particle is at rest, the rotation subgroup of the
Lorentz group will leave its four-momentum invariant.  This rotation
group will however change the direction of the particle spin.  We are
quite familiar with what the Pauli spin matrices do.  They are
the generators of the $SU(2)$ group which is locally isomorphic to
the three-dimensional rotation group.  In addition, the Pauli matrices
correspond the physical quantity known as the spin.  Indeed, Wigner's
work placed the spin of physics into Einstein's relativistic world.

It is not possible to bring a massless particle into its rest frame.
For this particle, we can can fix the coordinate system in which
the particle momentum is along the $z$ direction.  Wigner showed that
the little group for this particle is generated by the rotation around
the $z$ axis and two other generators consisting of linear combinations
of rotation and boost generators of the Lorentz group~\cite{wig39}.
He showed further that these three generators satisfy a closed set of
commutation relations, which is the same as the two-dimensional
Euclidean group consisting of one rotation and two translations.
It is not difficult to associate this rotation with the helicity of
the massless particle, but Wigner did not give a physical meaning to
those translation-like degrees of freedom.   It was not until 1971
that Janner and Janssen observed that they are gauge
transformations~\cite{janner71}.

There was another issue unresolved until the 1980s.  In
1953~\cite{inonu53}, Inonu and Wigner observed that the two-dimensional
Euclidean group can be obtained from a limiting case of the
three-dimensional rotation group.  This limiting case is known as
the group contraction.  The simplest way to grasp the picture is
to consider a sphere with a large radius.  It is then possible to
make a flat surface approximation on the north pole.  It is therefore
likely that the $E(2)$-like little group for a massless particle is
a limiting case of the $O(3)$-like little group.  It is also likely
that the limiting procedure is the infinite-momentum or zero-mass
limit.  But this problem was was not until 1990~\cite{kiwi90jm}.

Let us next see how this aspect of Wigner's work is relevant to
Feynman's parton picture.  Wigner's original paper~\cite{wig39}
tells us that the internal space-time symmetry groups for massive
and massless particles are different.  It took fifty years for the
physics community to establish that they are two different
manifestation of one covariant entity, as in the case of Einstein's
energy-momentum relation.  Likewise, we can raise the question of
whether the quark model and the parton model are two different
manifestation of one single covariant entity.

How can we approach this problem?  In their 1971 paper~\cite{fkr71},
Feynman {\it et al.} used harmonic oscillators to work out hadronic
mass spectra.  Even though they used relativistic oscillators, they
did not address the question of whether their formalism constitute a
representation of Wigner's little group.
On the other hand, Wigner did not use harmonic oscillators
too often in his papers, and Feynman did not pay enough attention to
Lorentz covariance.  Thus, in order to combine Feynman's initiative
with Wigner's formalism, we can construct representations of the
little group using harmonic oscillators.  In so doing, we construct
harmonic oscillator wave functions which can be Lorentz-boosted.

In Sec.~\ref{littleg}, we present a brief history of applications of
the little groups to internal space-time symmetries of relativistic
particles.  It is emphasized that massive and massless particles
have different internal space-time symmetries, namely the $O(3)$-like
and $E(2)$-like little groups respectively.
It is pointed out that the
translation-like transformations of the $E(2)$-like little group
corresponds to gauge transformations.
In Sec.~\ref{contrac}, it is shown that the $E(2)$-like little group
can be obtained as an infinite-momentum or zero-mass limit of
the $O(3)$-like little group for a massive particle.  This procedure
is like the the original Inonu-Wigner contraction in which $O(3)$
becomes contracted into $E(2)$.

In Sec.~\ref{feynm}, it is noted that R. P. Feynman was quite fond of
harmonic oscillators, especially trying new physical ideas.
It is pointed out that harmonic oscillators embrace many
useful properties of the Lorentz group.
We show in Sec.~\ref{par} that this $O(1,1)$ formalism enables to
construct a covariant model of relativistic extended particles.  As a
consequence, we show that the quark and parton model are two different
aspects of one covariant object.  It is shown also that this parton
picture exhibits the decoherence effect.

\section{Wigner's Spins}\label{littleg}

The group of Lorentz transformations consists of three boosts and
three rotations.  The rotations therefore constitute a subgroup of
the Lorentz group.  If a massive particle is at rest, its four-momentum
is invariant under rotations.  Thus the little group for a massive
particle at rest is the three-dimensional rotation group.  Then what is
affected by the rotation?  The answer to this question is very simple.
The particle in general has its spin.  The spin orientation is going
to be affected by the rotation!

If the rest-particle is boosted along the $z$ direction, it will pick
up a non-zero momentum component.  The generators of the $O(3)$ group
will then be boosted.  The boost will take the form of conjugation by
the boost operator.  This boost will not change the Lie algebra of the
rotation group, and the boosted little group will still leave the
boosted four-momentum invariant.  We call this the $O(3)$-like little
group.

We use in this report the metric convention $(x, y, z, t)$.  Then
the Lorentz group is generated by the three rotation generators
$J_{i}$ and three boost generators $K_{i}$ whose third components
take the form
\begin{equation}\label{j3}
J_{3} = \pmatrix{0 & -i & 0 & 0 \cr i & 0 & 0 & 0
\cr 0 & 0 & 0 & 0 \cr 0 & 0 & 0 & 0} ,  \qquad
K_{3} = \pmatrix{0 & 0 & i & 0 \cr 0 & 0 & 0 & 0
\cr 0 & 0 & 0 & 0 \cr 0 & 0 & i & 0} ,
\end{equation}
respectively.  The expressions for the first and second components
are well known.

It is important to note that the rotation generators satisfy the
closed set of commutation relations:
\begin{equation}
 [J_{i}, J_{j}] = i\epsilon_{ijk} J_{k} ,
\end{equation}
but the commutation relations among the $K_{i}$ operators do not.
They lead to
\begin{equation}
 [K_{i}, K_{j}] = -i\epsilon_{ijk} J_{k} .
\end{equation}
However, we can still consider the following combinations of the
generators.
\begin{equation}\label{n1n2}
N_{1} = K_{1} - J_{2} , \qquad N_{2} = K_{2} + J_{1} ,
\end{equation}
Then, together with $J_{3}$, they form the following closed set of
commutation relations.
\begin{equation}\label{e2lcom}
[N_{1}, N_{2}] = 0 , \qquad [J_{3}, N_{1}] = iN_{2} , \qquad
[J_{3}, N_{2}] = -iN_{1} .
\end{equation}

It is not difficult to associate the rotation generator $J_{3}$ with
the helicity degree of freedom of the massless particle.   Then what
physical variable is associated with the $N_{1}$ and $N_{2}$ generators?
Indeed, Wigner was the one who discovered the existence of these
generators, but did not give any physical interpretation to these
translation-like generators.  For this reason, for many years, only
those representations with the zero-eigenvalues of the $N$ operators
were thought to be physically meaningful representations~\cite{wein64}.

It is possible to get a hint that the $N$ operators generate gauge
transformations from Weinberg's
1964 papers~\cite{wein64,hks82}.  But it was not until 1971 when
Janner and Janssen explicitly demonstrated that they generate gauge
transformations~\cite{janner71,kuper76}.  In order to fully appreciate
their work, let us compute the transformation matrix
\begin{equation}
 \exp{(-i(uN_{1} + vN_{2})}
\end{equation}
generated by $N_{1}$ and $N_{2}$.  Then the four-by-four matrix takes
the form
\begin{equation}\label{trans}
\pmatrix{1 & 0 &-u & u \cr 0 & 1 & -v & v \cr
u & v & 1 - (u^{2} + v^{2})/2 & (u^{2} + v^{2})/2 \cr
u & 0 & -(u^{2} + v^{2})/2 & 1 + (u^{2} + v^{2})/2} .
\end{equation}
If we apply this matrix to the four-vector to the four-momentum vector
\begin{equation}\label{4mom}
 p = (0, 0, \omega, \omega)
\end{equation}
of a massless particle, the momentum remains invariant.  It therefore
satisfies the condition for the little group.  If we apply this matrix
to the electromagnetic four-potential
\begin{equation}
 A = (A_{1}, A_{2}, A_{3}, A_{0}) \exp{(i(kz -\omega t))} ,
\end{equation}
with $A_{3} = A_{0}$ which is the Lorentz condition,
the result is a gauge transformation.  This is what Janner and Janssen
discovered in their 1971 and 1972 papers~\cite{janner71}. Thus
the matrices $N_{1}$ and $N_{2}$ generate gauge transformations.

\section{The E(2)-like Little as a Contraction of the O(3)-like
    Little Group}\label{contrac}

We know how to boost the the four-momentum of the rest particle
along the $z$ direction.  The boost matrix takes the form
\begin{equation}\label{simil2}
 \pmatrix{1&0&0&0\cr0&1&0&0\cr0&0 & \cosh\eta & \sinh\eta
\cr0 & 0 & \sinh\eta & \cosh\eta} ,
\end{equation}
Then the the generators of the $O(3)$-like little group can also
be boosted.  Under this boost, $J_{3}$ will remain invariant, but
$J_{1}$ and $J_{2}$ will become
\begin{equation}
J'_{1} = B~J_{1}~B^{-1}, \qquad J'_{2} = B~J_{2}~B^{-1} ,
\end{equation}
Because the above transformations are similarity transformations,
these new generators will still satisfy the commutation relations
for the three-dimensional rotation group.

On the other hand, we can take the following limit in the spirit
of the Inonu-Wigner group contraction.
\begin{equation}
 N_{1} = {1\over R} B^{-1} J_{2} B ,
\qquad N_{2} = -{1\over R} B^{-1} J_{1} B ,
\end{equation}
where $N_{1}$ and $N_{2}$ are given in Eq.(\ref{n1n2}).  The generators
$N_{1}$ and $N_{2}$ are the contracted $J_{2}$ and $J_{1}$ respectively
in the infinite-momentum and/or zero-mass limit~\cite{kiwi90jm}.
It was noted that $N_{1}$ and $N_{2}$ generate
gauge transformations on massless particles.  Thus the contraction of
the transverse rotations leads to gauge transformations.

We have seen in this section that Wigner's $O(3)$-like little group
can be contracted into the $E(2)$-like little group for massless
particles.  Here, we worked out explicitly for the spin-1 case, but
this mechanism should be applicable to all other spins.  Of particular
interest is spin-1/2 particles.  This has been studied by Han, Kim
and Son~\cite{hks82}.  They noted that there are also gauge
transformations for spin-1/2 particles, and the polarization of
neutrinos is a consequence of gauge invariance.  It has also been
shown that the gauge dependence of spin-1 particles can be traced to
the gauge variable of the spin-1/2 particle~\cite{hks86jm}.  It would
be very interesting to see how the present formalism is applicable to
higher-spin particles.

Another case of interest is the space-time symmetry of relativistic
extended particles.  In 1973~\cite{kn73}, Kim and Noz constructed
a ground-state harmonic oscillator wave function which can be
Lorentz-boosted.  It was later found that this oscillator formalism
can be extended to represent the $O(3)$-like little
group~\cite{kno79,knp86}.  This oscillator formalism has a stormy
history because it ultimately plays a pivotal role in combining
quantum mechanics and special relativity~\cite{dir45,yuka53}.

With these wave functions, we propose to solve the following problem in
high-energy physics.  The quark model works well when hadrons are at
rest or move slowly.  However, when they move with speed close to that
of light, they appear as a collection of an infinite-number of
partons~\cite{fey69}.  The question then is whether the parton model is
a Lorentz-boosted quark model.  This question has been addressed
before~\cite{kn77par,kim89}, but it can generate more interesting
problems~\cite{kiwi90pl,hkn99ajp}.  The present situation is presented
in the Table~\ref{table1}.

\begin{table}[thb]

\caption{Massive and massless particles in one package.  Wigner's
little group unifies the internal space-time symmetries for massive and
massless particles.  It is a great challenge for us to find
another unification: the unification of the quark and parton pictures in
high-energy physics.}\label{table1}

\vspace{2mm}
\small

\begin{center}

\begin{tabular}{lccc}

\hline
{}&{}&{}&{}\\
{} & Massive, Slow \hspace{6mm} & COVARIANCE \hspace{6mm}&
Massless, Fast \\[4mm]\hline
{}&{}&{}&{}\\
Energy- & {}  & Einstein's & {} \\
Momentum & $E = p^{2}/2m$ & $ E = [p^{2} + m^{2}]^{1/2}$ & $E = p$
\\[4mm]\hline
{}&{}&{}&{}\\
Internal & $S_{3}$ & {}  &  $S_{3}$ \\[-1mm]
Space-time &{} & Wigner's  & {} \\ [-1mm]
Symmetry & $S_{1}, S_{2}$ & Little Group & Gauge Trans. \\[4mm]\hline
{}&{}&{}&{}\\
Relativistic & {} & One  &  {} \\[-1mm]
Extended & Quark Model & Covariant  & Parton Model\\ [-1mm]
Particles & {} & Theory &{} {} \\[4mm]\hline

\end{tabular}

\end{center}
\end{table}


We are now ready to consider the third row of Table~\ref{table1}.
In the this table, we would like to say that the quark model
and the parton model are two different manifestation of one covariant
entity.  In order to appreciate fully this covariant aspect, let us
examine Feynman's style of doing physics.

\section{Feynman's Harmonic Oscillators}\label{feynm}

Quantum field theory has been quite successful in terms of
perturbation techniques in quantum electrodynamics.  However, this
formalism is basically based
on the S matrix for scattering problems and useful only for physically
processes where a set of free particles becomes another set of free
particles after interaction.  Quantum field theory does not address
the question of localized probability distributions and their
covariance under Lorentz transformations.
The Schr\"odinger quantum mechanics of the hydrogen atom deals with
localized probability distribution.  Indeed, the localization condition
leads to the discrete energy spectrum.  Here, the uncertainty relation
is stated in terms of the spatial separation between the proton and
the electron.  If we believe in Lorentz covariance, there must also
be a time separation between the two constituent particles.

Let us use the simplest hadron consisting of
two quarks bound together with an attractive force, and consider their
space-time positions $x_{a}$ and $x_{b}$, and use the variables
\begin{equation}
X = (x_{a} + x_{b})/2 , \qquad x = (x_{a} - x_{b})/2\sqrt{2} .
\end{equation}
The four-vector $X$ specifies where the hadron is located in space and
time, while the variable $x$ measures the space-time separation
between the quarks.  According to Einstein, this space-time separation
contains a time-like component which actively participates in the Lorentz
transformation, as can be
seen from
\begin{equation}\label{boostm}
\pmatrix{z' \cr t'} = \pmatrix{\cosh \eta & \sinh \eta \cr
\sinh \eta & \cosh \eta } \pmatrix{z \cr t} ,
\end{equation}
when the hadron is boosted along the $z$ direction.
In terms of the light-cone variables defined as~\cite{dir49}
\begin{equation}
u = (z + t)/\sqrt{2} , \qquad v = (z - t)/\sqrt{2} .
\end{equation}
The boost transformation of Eq.(\ref{boostm}) takes the form
\begin{equation}\label{lorensq}
u' = e^{\eta } u , \qquad v' = e^{-\eta } v .
\end{equation}
The $u$ variable becomes expanded while the $v$ variable becomes
contracted.

\begin{figure}
\centerline{\psfig{figure=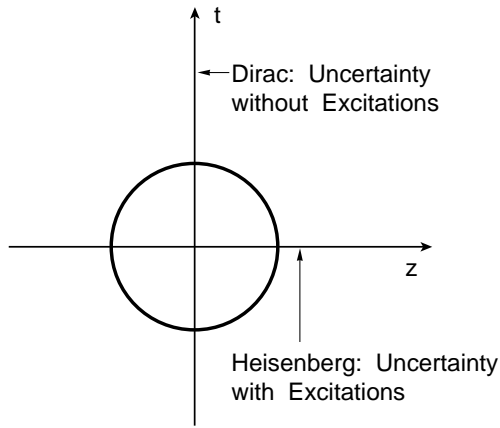,angle=0,width=70mm}}
\vspace{0mm}
\caption{Space-time picture of quantum mechanics.  There
are quantum excitations along the space-like longitudinal direction, but
there are no excitations along the time-like direction.  The time-energy
relation is a c-number uncertainty relation.}\label{quantum}
\end{figure}

Does this time-separation variable exist when the hadron is at rest?
Yes, according to Einstein.  In the present form of quantum mechanics,
we pretend not to know anything about this variable.
There is also an uncertainty
relation between the time and energy variables.  However, there are
no known time-like excitations.  Unlike Heisenberg's
uncertainty relation applicable to position and momentum, the time and
energy separation variables are c-numbers, and we are not allowed to
write down the commutation relation between them.  Indeed, the
time-energy uncertainty relation is a c-number uncertainty
relation~\cite{dir27}, as is illustrated in Fig.~\ref{quantum}.

How does this space-time asymmetry fit into the world of Lorentz
covariance~\cite{kn73}?  The
answer is that Wigner's $O(3)$-like little group is not a
Lorentz-invariant symmetry, but is a covariant symmetry~\cite{wig39}.
It has been shown that the time-energy uncertainty applicable to the
time-separation variable fits perfectly into the $O(3)$-like symmetry
of massive relativistic particles~\cite{knp86}.

\begin{figure}
\centerline{\psfig{figure=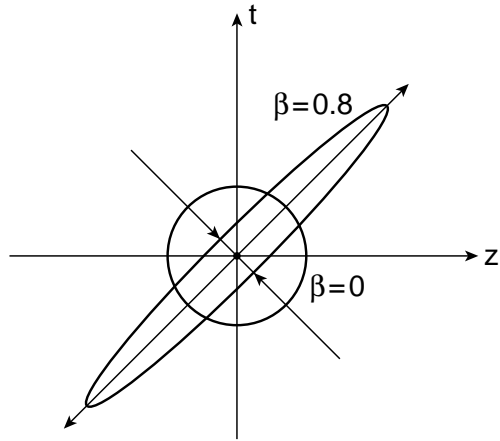,angle=0,width=70mm}}
\caption{Effect of the Lorentz boost on the space-time
wave function.  The circular space-time distribution at the rest frame
becomes Lorentz-squeezed to become an elliptic
distribution.}\label{ellipse}
\end{figure}

The c-number time-energy uncertainty relation allows us to write down
a time distribution function without excitations~\cite{knp86}.
If we use Gaussian forms for both space and time distributions, we
can start with the expression
\begin{equation}\label{ground}
\left({1 \over \pi} \right)^{1/2}
\exp{\left\{-{1 \over 2}\left(z^{2} + t^{2}\right)\right\}}
\end{equation}
for the ground-state wave function.
In their classic 1971 paper~\cite{fkr71}, Feynman {\it et al.} start
with the following Lorentz-invariant differential equation.
\begin{equation}\label{osceq}
{1\over 2} \left\{x^{2}_{\mu} -
{\partial^{2} \over \partial x_{\mu }^{2}}
\right\} \psi(x) = \lambda \psi(x) .
\end{equation}
This partial differential equation has many different solutions
depending on the choice of separable variables and boundary conditions.
Feynman {\it et al.} insist on Lorentz-invariant solutions which are
not normalizable.  On the other hand, if we insist on normalization,
the ground-state wave function takes the form of Eq.(\ref{ground}).
It is then possible to construct a representation of the
Poincar\'e group from the solutions of the above differential
equation~\cite{knp86}.  If the system is boosted, the wave function
becomes
\begin{equation}\label{eta}
\psi_{\eta }(z,t) = \left({1 \over \pi }\right)^{1/2}
\exp\left\{-{1\over 2}\left(e^{-2\eta }u^{2} +
e^{2\eta}v^{2}\right)\right\} .
\end{equation}
This wave function becomes Eq.(\ref{ground}) if $\eta$ becomes zero.
The transition from Eq.(\ref{ground}) to Eq.(\ref{eta}) is a
squeeze transformation.  The wave function of Eq.(\ref{ground}) is
distributed within a circular region in the $u v$ plane, and thus
in the $z t$ plane.  On the other hand, the wave function of
Eq.(\ref{eta}) is distributed in an elliptic region with the light-cone
axes as the major and minor axes respectively, as illustrated in
Fig.~\ref{ellipse}.  If $\eta$ becomes very
large, the wave function becomes concentrated along one of the
light-cone axes.  Indeed, the form given in Eq.(\ref{eta}) is a
Lorentz-squeezed wave  function.


\section{Feynman's Parton Picture}\label{par}
In 1969, Feynman observed that a fast-moving hadron
can be regarded as a collection of many ``partons'' whose properties
do not appear to be quite different from those of the
quarks~\cite{fey69}.  For example, the number of quarks inside a
static proton is three, while the number of partons in a rapidly
moving proton appears to be infinite.  The question then is how
the proton looking like a bound state of quarks to one observer
can appear different to an observer in a different Lorentz frame?
Feynman made the following systematic observations.

\begin{itemize}

\item[a.]  The picture is valid only for hadrons moving with
  velocity close to that of light.

\item[b.]  The interaction time between the quarks becomes dilated,
   and partons behave as free independent particles.

\item[c.]  The momentum distribution of partons becomes widespread as
   the hadron moves fast.

\item[d.]  The number of partons seems to be infinite or much larger
    than that of quarks.

\end{itemize}

\noindent Because the hadron is believed to be a bound state of two
or three quarks, each of the above phenomena appears as a paradox,
particularly b) and c) together.

\begin{figure}
\centerline{\psfig{figure=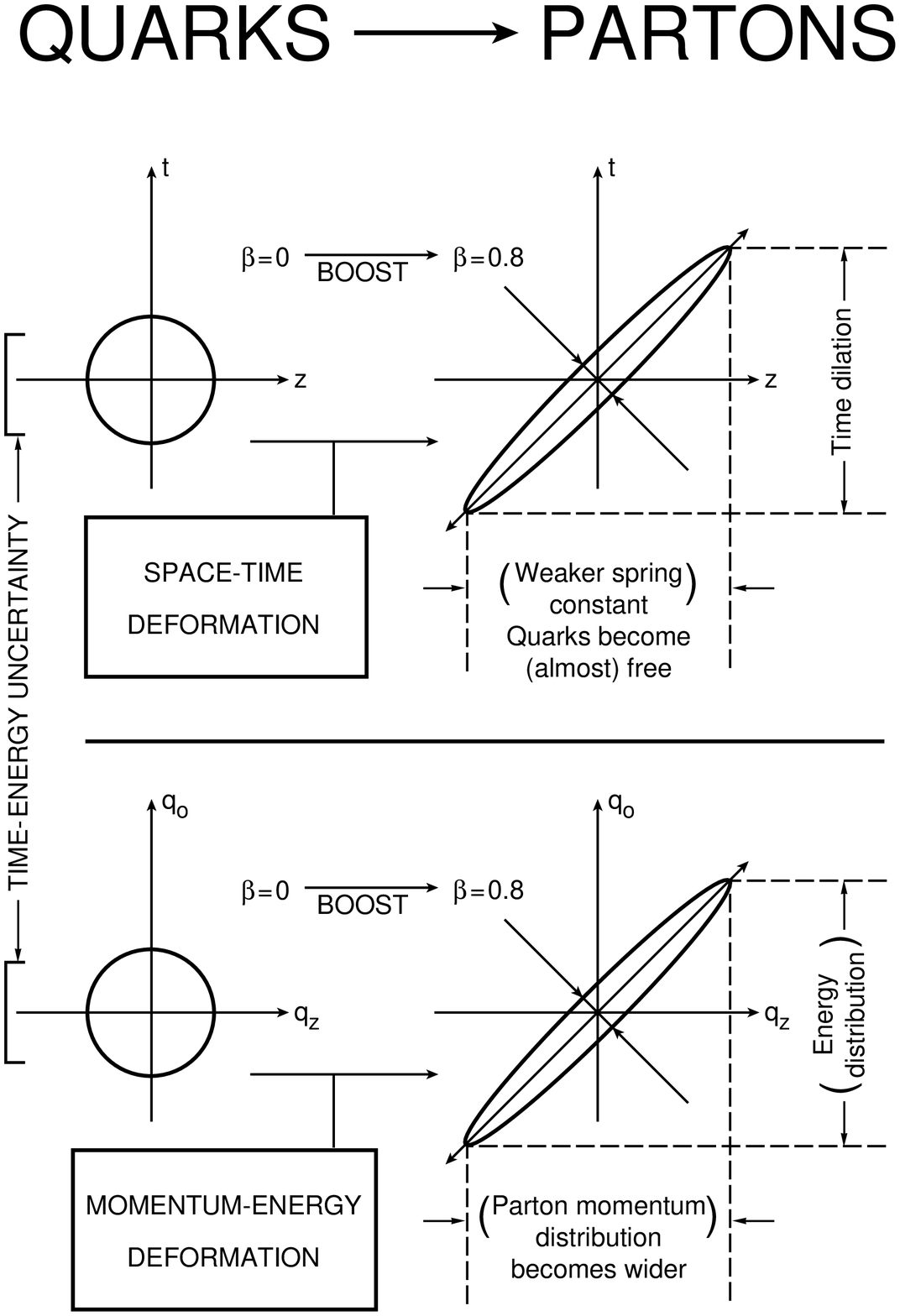,angle=0,width=100mm}}
\vspace{2mm}
\caption{Lorentz-squeezed space-time and momentum-energy
wave functions.  As the hadron's speed approaches that of light, both
wave functions become concentrated along their respective positive
light-cone axes.  These light-cone concentrations lead to Feynman's
 parton picture.}\label{parton}
\end{figure}
In order to resolve this paradox, let us write down the
momentum-energy wave function corresponding to Eq.(\ref{eta}).
If the quarks have the four-momenta $p_{a}$ and $p_{b}$, we can
construct two independent four-momentum variables~\cite{fkr71}
\begin{equation}
P = p_{a} + p_{b} , \qquad q = \sqrt{2}(p_{a} - p_{b}) ,
\end{equation}
where $P$ is the total four-momentum and is thus the hadronic
four-momentum.  $q$ measures the four-momentum separation between
the quarks.  Their light-cone variables are
\begin{equation}\label{conju}
q_{u} = (q_{0} - q_{z})/\sqrt{2} ,  \qquad
q_{v} = (q_{0} + q_{z})/\sqrt{2} .
\end{equation}
The resulting momentum-energy wave function is
\begin{equation}\label{phi}
\phi_{\eta }(q_{z},q_{0}) = \left({1 \over \pi }\right)^{1/2}
\exp\left\{-{1\over 2}\left(e^{-2\eta}q_{u}^{2} +
e^{2\eta}q_{v}^{2}\right)\right\} .
\end{equation}
Because we are using here the harmonic oscillator, the mathematical
form of the above momentum-energy wave function is identical to that
of the space-time wave function.  The Lorentz squeeze properties of
these wave functions are also the same.  This aspect of the squeeze
has been exhaustively discussed in the
literature~\cite{knp86,kn77par,kim89}.

When the hadron is at rest with $\eta = 0$, both wave functions
behave like those for the static bound state of quarks.  As $\eta$
increases, the wave functions become continuously squeezed until
they become concentrated along their respective positive
light-cone axes.  Let us look at the z-axis projection of the
space-time wave function.  Indeed, the width of the quark distribution
increases as the hadronic speed approaches that of the speed of
light.  The position of each quark appears widespread to the observer
in the laboratory frame, and the quarks appear like free particles.

The momentum-energy wave function is just like the space-time wave
function, as is shown in Fig.~\ref{parton}.  The longitudinal momentum
distribution becomes wide-spread as the hadronic speed approaches the
velocity of light.  This is in contradiction with our expectation from
nonrelativistic quantum mechanics that the width of the momentum
distribution is inversely proportional to that of the position wave
function.  Our expectation is that if the quarks are free, they must
have their sharply defined momenta, not a wide-spread distribution.

However, according to our Lorentz-squeezed space-time and
momentum-energy wave functions, the space-time width and the
momentum-energy width increase in the same direction as the hadron
is boosted.  This is of course an effect of Lorentz covariance.
This indeed is the key to the resolution of the quark-parton
paradox~\cite{knp86,kn77par}.

The most puzzling problem in the parton picture is that partons in
the hadron appear as incoherent particles, while quarks are coherent
when the hadron is at rest.  Does this mean that the coherence is
destroyed by the Lorentz boost?   The answer is NO, and here is the
resolution to this puzzle.

When the hadron is boosted, the hadronic matter becomes squeezed and
becomes concentrated in the elliptic region along the positive
light-cone axis  The length of the major axis becomes expanded by
$e^{\eta}$, and the minor axis is contracted by $e^{-\eta}$.

This means that the interaction time of the quarks among themselves
become dilated.  Because the wave function becomes wide-spread, the
distance between one end of the harmonic oscillator well and the
other end increases.  This effect, first noted by Feynman~\cite{fey69},
is universally observed in high-energy hadronic experiments.  The
period is oscillation increases like $e^{\eta}$.

On the other hand, the interaction time with the external signal,
since it is moving in the direction opposite to the direction of
the hadron, travels along the negative light-cone axis.
If the hadron contracts along the negative light-cone axis, the
interaction time decreases by $e^{-\eta}$.  The ratio of the interaction
time to the oscillator period becomes $e^{-2\eta}$.  The energy of each
proton coming out of the Fermilab accelerator is $900 GeV$.  This leads
the ratio to $10^{-6}$.  This is indeed a small number.  The external
signal is not able to sense the interaction of the quarks among
themselves inside the hadron.

This covariant picture of Feynman's parton model can be placed on the
third row of Table~\ref{table1}.

\bigskip
{\small I would like to thank Professor Miroslav Finger and the Local
Organizing Committee for inviting me to this Advanced Summer Institute
and to the City of Prague.}

\end{document}